\documentclass[11pt]{article}
\usepackage{indentfirst}

\usepackage{graphics,graphicx}
\usepackage{indentfirst}
\usepackage{makeidx}
\usepackage{amssymb,amsmath}
\usepackage{wasysym}
\usepackage[usenames, dvipsnames]{color}
\usepackage[ pdftex, plainpages = false, pdfpagelabels,
                 pdfpagelayout = useoutlines,
                 bookmarks,
                 bookmarksopen = true,
                 bookmarksnumbered = true,
                 breaklinks = true,
                 linktocpage=all,
                 pagebackref=false,
                 colorlinks = true,
                 linkcolor = BrickRed,
                 urlcolor  = blue,
                 citecolor = BrickRed,
                 anchorcolor = green,
                 hyperindex = true,
                 hyperfigures
                 ]{hyperref}
\usepackage{supertabular}
\usepackage{xifthen} 

\usepackage{etex}

\newcommand{\bv}{\begin{verse}}
\newcommand{\ev}{\end{verse}}
\newcommand{\be}{\begin{equation}}
\newcommand{\ee}{\end{equation}}
\newcommand{\bea}{\begin{eqnarray}}
\newcommand{\eea}{\end{eqnarray}}
\newcommand{\bq}{\begin{quotation}}
\newcommand{\eq}{\end{quotation}}

\newcommand{\myurl}[2][]{\ifthenelse{\isempty{#1}}{\url{#2}}{\href{#1}{\tt #2}}}

\newtheorem{cabello}{Ad\'anism}
\newcommand{\bac}{\begin{cabello}\protect$\!\!${\em\bf :}$\;\;$}
\newcommand{\eac}{\end{cabello}}

\newtheorem{weinberg}{Weinbergism}
\newcommand{\bsw}{\begin{weinberg}\protect$\!\!${\em\bf :}$\;\;$}
\newcommand{\esw}{\end{weinberg}}

\begin{document}

\begin{center}
\LARGE {\bf On Participatory Realism}\bigskip\bigskip\\
\Large Christopher A. Fuchs \bigskip \\
\normalsize Department of Physics, University of Massachusetts Boston \\ 100 Morrissey Boulevard, Boston MA 02125, USA \smallskip \\
and \smallskip \\
Max Planck Institute for Quantum Optics \\ Hans-Kopfermann-Strasse 1, 85748 Garching, Germany
\bigskip\\
\large 12 January 2016
\end{center}

\bigskip\bigskip

\bq
\noindent
{\bf Abstract:}  In the {\sl Philosophical Investigations}, Ludwig Wittgenstein wrote, ``\,`I' is not the name of a person, nor `here' of a place, \ldots.  But they are connected with names. \ldots\ \  [And] it is characteristic of physics not to use these words.''  This statement expresses the dominant way of thinking in physics:  Physics is about the impersonal laws of nature; the ``I'' never makes an appearance in it.  Since the advent of quantum theory, however, there has always been a nagging pressure to insert a first-person perspective into the heart of physics.  In incarnations of lesser or greater strength, one may consider the ``Copenhagen'' views of Bohr, Heisenberg, and Pauli, the observer-participator view of John Wheeler, the informational interpretation of Anton Zeilinger and \v{C}aslav Brukner, the relational interpretation of Carlo Rovelli, and, most radically, the QBism of N. David Mermin, R\"udiger Schack, and the present author, as acceding to the pressure.  These views have lately been termed ``participatory realism'' to emphasize that rather than relinquishing the idea of reality (as they are often accused of), they are saying that reality is {\it more\/} than any third-person perspective can capture.  Thus, far from instances of instrumentalism or antirealism, these views of quantum theory should be regarded as attempts to make a deep statement about the nature of reality.  This paper explicates the idea for the case of QBism.  As well, it highlights the influence of John Wheeler's ``law without law'' on QBism's formulation.
\eq

\bigskip

John Archibald Wheeler's writings were a tremendous influence on the point of view on quantum mechanics we now call QBism \cite{Fuchs10a}, developed by the author \cite{Fuchs02a, Fuchs02b, Fuchs07a, Fuchs11a} and colleagues Carlton Caves \cite{Caves02a, Caves02b, Caves02c, Caves07}, Asher Peres \cite{Fuchs00a,Fuchs00b}, Marcus Appleby \cite{Appleby05a,Appleby05b,Appleby05c,Appleby09a,Appleby06a,Appleby13a,Appleby15a}, Blake Stacey \cite{Stacey14a,Stacey14b,Stacey15a}, Hans Christian von Baeyer \cite{vonBaeyer16a}, David Mermin \cite{Mermin09a, Mermin09b, Mermin12a, Mermin12b, Mermin13a, Mermin13b, Mermin14a, Mermin14b, Mermin14c, Mermin14d, Fuchs14a, Mermin14f, Fuchs15a}, and most especially R\"udiger Schack \cite{Schack01a, Fuchs04a, Fuchs04b, Fuchs09a, Fuchs11b, Fuchs12a, Fuchs13a, Fuchs14b, Fuchs15b}.\footnote{A further important review of QBism by an outsider can be found in \cite{Timpson08a}.}  It is true that the term was initially an abbreviation for Quantum Bayesianism, but long before Bayesianism was brought into the mix of my own thinking (I won't speak for the others), there were John Wheeler's exhortations on ``law without law'' filling my every sleepless night.  A typical something Wheeler would write on the subject would sound like this,
\bq\noindent
``Law without law'': It is difficult to see what else than that can be the ``plan'' for physics. It is preposterous to think of the laws of physics as installed by a Swiss watchmaker to endure from everlasting to everlasting when we know that the universe began with a big bang. The laws must have come into being.  Therefore they could not have been always a hundred percent accurate.  That means that they are derivative, not primary.  Also derivative, also not primary is the statistical law of distribution of the molecules of a dilute gas between two intersecting portions of a total volume. \ This law is always violated and yet always upheld.  The individual molecules laugh at it; yet as they laugh they find themselves obeying it. \ldots\ \ Are the laws of physics of a similar statistical character?  And if so, statistics of what?  Of billions and billions of acts of observer-participancy which individually defy all law? \ldots\ \ [Might] the entirety of existence, rather than [be] built on particles or fields or multidimensional geometry, [be] built on billions upon billions of elementary quantum phenomena, those elementary acts of observer-participancy?
\eq
Roger Penrose called the idea ``barely credible,''\cite{Penrose90} but there was something about it that attracted me like nothing else in physics:  Right or wrong---or more likely, not well-defined enough to be {\it right\/} or {\it wrong}!---Wheeler's thinking was the very reason I pursued the study of physics in the first place.
In college and graduate school, I would read every piece of Wheeler's writing on the subject that I could find, no matter how repetitious the task became.\footnote{The exact corpus of my readings was this: \cite{Wheeler71a, Wheeler73a, Wheeler74a, Wheeler74b, Wheeler75a, Wheeler75b, Wheeler75c, PattonWheeler75, Wheeler76a, WheelerPatton77, Wheeler77a, Wheeler77b, Wheeler78a, Wheeler79a, Wheeler79b, Wheeler79c, Wheeler79d, Wheeler79e, Wheeler80a, Wheeler80b, Wheeler80c, Wheeler80d, Wheeler81a, Wheeler81b, Wheeler81c, Wheeler82a, Wheeler82b, Wheeler82c, Wheeler82d, Wheeler82e, Wheeler83a, Wheeler83b, Wheeler83c, Wheeler83d, Wheeler83e, Wheeler84a, Wheeler84c, Wheeler84d, Wheeler85a, Wheeler85b, Wheeler85c, Wheeler86a, Wheeler86b, Wheeler86c, Wheeler86d, Wheeler87a, Wheeler87b, Wheeler88a, Wheeler88b, Wheeler88c, Wheeler89a, Wheeler90a, Wheeler92a, Wheeler92b, Wheeler94a, Wheeler98a, Wheeler98b, Wheeler98c, Wheeler99a, Wheeler00}.}

One thing that is absolutely clear in Wheeler's writings is that the last thing he was pursuing was an {\it instrumentalist\/} understanding of quantum theory.  He thought that quantum theory was the deepest statement on {\it nature\/} and {\it reality\/} ever discovered by physics.  Thus, he would play little dialogues with himself like this one: ``The strange necessity of the quantum as we see it everywhere in the scheme of physics comes from the requirement that---via observer-participancy---the universe should have a way to come into being?  If true---and it is attractive---it should provide someday a means to \underline{derive} quantum mechanics from the requirement that the universe must have a way to come into being.''  Indeed, he was fond of quoting Niels Bohr, who in his last interview (the day before he died), said, ``I felt \ldots\ that philosophers were very odd people who really were lost, because they have not the instinct that it is important to learn something and that we must be prepared really to learn something of very great importance \ldots .''  Instrumentalists don't learn lessons about nature!  They just try to predict what is coming next as best they can.

So, it is with some dismay and consternation that I hear over and over that QBism is an instrumentalist account of quantum theory.  For instance, in Jeffrey Bub's recent book {\sl Bananaworld:\ Quantum Mechanics for Primates} \cite{Bub16a}\footnote{But other similar accounts abound.  Here is another choice one, from Ladyman and Ross's book {\sl Everything Must Go:\ Metaphysics Naturalized} \cite{Ladyman07a}:  ``According to [Fuchs and Peres] the quantum state of a system is just a probability distribution in disguise. This is an instrumentalist approach that is metaphysically unhelpful.''}, he writes,
\bq
Another approach to the conceptual problems of quantum mechanics is the ``quantum Bayesianism'' or ``QBism'' of Christopher Fuchs and R\"udiger Schack.  David Mermin is a recent convert.  \ldots\ \ On this view, all probabilities, are understood in the subjective sense as the personal judgements of an agent, based on how the external world responds to an action by the agent.  Then there's no problem in explaining how quantum probabilities of ``what you'll obtain if you measure'' are related to classical ignorance of ``what's there,'' because all probabilities are personal bets about what you'll find if you look.  This approach seems to be straightforwardly instrumentalist from the start:  a theory is a useful tool, a predictive device, based on a summary of how the world responds to an agent's actions.
\eq
When I read things like this, I find myself wanting to say, ``[P]hilosophers [are] very odd people who really [are] lost, because they have not the instinct that it is important to learn something and that we must be prepared really to learn something of very great importance \ldots''  What is at stake with quantum theory is the very nature of reality. Should reality be understood as something completely impervious to our interventions, or should it be viewed as something responsive to the very existence of human beings?  Was the big bang all and the whole of creation?  Or is creation going on all around us all the time, and we ourselves are taking part in it?  No philosophical predisposition to instrumentalism led inexorably to QBism:  Quantum theory itself threw these considerations before us!

The bulk of this paper is made of correspondence, in the vein of a couple of my previous contributions to the literature \cite{Samizdat1,Samizdat2}, where I try to set the record straight on this.  This correspondence was prompted predominantly by some early drafts of Ad\'an Cabello's paper, ``Interpretations of Quantum Theory:\ A Map of Madness,'' \cite{Cabello15a} the final form of which is posted here: \myurl{http://arxiv.org/abs/1509.04711}.  Initially Ad\'an labelled the views on quantum mechanics of the Copenhagen School, \v{C}aslav Brukner and Anton Zeilinger, Asher Peres, Carlo Rovelli, John Wheeler, and the QBism of David Mermin, R\"udiger Schack, and myself, as all ``non-realist.'' [See Ad\'an's paper for appropriate citations to these various views.] This incensed me so, bringing back waves of emotion over having QBism called instrumentalism, solipsism, anti-realism, mysticism, and even psychology, that I urged the term ``participatory realism'' to try to capture what is in common among these views, while calling attention outright that they are all {\it realist\/} takes on the task of physics.  It also seemed a worthy tribute to John Wheeler, as I thought he captured the appropriate sentiment with his phrase ``participatory universe''---it's as full-blown a notion of reality as anyone could want, recognizing only that the users of quantum mechanics have their part of it too.  That's not less reality, that's more.  Ultimately Ad\'an adopted the term ``Type-II (participatory realism)'' for his paper.

After my note of 28 July 2015 (Section 2 below), Ad\'an responded with,
\bac
There is a part of your e-mail that I found deeply illuminating.  It is the part ending in:
\bq
\noindent {\rm QBism takes the idea of irreducible randomness much further than Zeilinger or Brukner.  Quantum theory advises us to recognize that no matter how we slice up nature, we will never find any pieces of it beholden to {\bf laws} in the usual physics sense of the term.  John Wheeler toyed with the idea that ``the only law of nature is that there is no law,'' and quantum theory on a QBist reading supports this.}
\eq
THIS IS IT\@!  I LOVE IT!

It may seem that you are denying the real or simply being elusive about the real, while you are actually making a strong statement about the real.
\eac

I was very happy to see how this realization finally clicked for Ad\'an.  I hope it will the same for you, my new reader.

\section{``QBism as Realism,'' to Steve Weinberg, 7 August 2015}

\bsw
When you say that you are averse to calling the Born rule a law of nature, ``at least in the sense that you seem to support in some of your correspondence with David [Mermin]'' I take this simply to mean that you do not take seriously the existence of any laws of nature of the sort I hope for, laws that are part of objective reality, independent of who is studying them or if anyone is studying them.  I don't want to attribute to you or David or QBism any view one way or another about the laws of nature, but I have inserted a new short paragraph at the top of p.\ 8 in the attached version to try to clarify the issue in general terms.
\esw

I have just spent a little while reviewing what you had written on QBism in Chapter 3 of your book\footnote{Steven Weinberg, {\sl Lectures on Quantum Mechanics}, 2nd Edition, (Cambridge U Press, Cambridge, UK, 2015).} and in some of our correspondence.  I much appreciate your effort to not ``misrepresent'' us, as you put it.  In my last iteration to you, I didn't make a big stink of your calling QBism a form of  ``instrumentalism'' because I felt it would be too much of a burden on you at the 11th hour.  Also I knew that there would be plenty of subtleties involved in our coming to a mutual understanding of what each of us means by so simple a phrase as ``laws of nature.''  In other words, it just didn't seem worth it to have a discussion.

In the meantime, however, I have put some effort into expressing more clearly the sense in which I see QBism as fitting into a realist project for physics.  It was in response to a chart Ad\'an Cabello of Seville has been putting together to try to capture the distinctions in various interpretations as regards their ``assessment of reality.''  In that chart he had initially labelled QBism as ``non-realist.''  For completeness, and so you'll understand the context of this note, I'll attach Ad\'an's draft, titled ``Map of Madness.''

Below are a couple of responses I sent him, which I spent some time and care in composing.  I hope you will find them clarifying too.  At the end, Ad\'an wrote me, ``It may seem that you are denying the real or simply being elusive about the real, while you are actually making a strong statement about the real.''  That indeed is what I like to think of QBism---very much taking it out of the camp of instrumentalism---but you may well continue to disagree.

By the way, one of the pieces below includes some paragraphs that were cut from a Q\&A at {\sl Quanta Magazine}, which I see you have also participated in.  I wish my photo there were so classy as yours!

\section{``Realisms,'' to Ad\'an Cabello, 28 July 2015}

I've owed you a note for some time now.  Let me finally sit down and try to devote the time to respond properly.  Sorry to keep you waiting.

Realisms.  The first thing I will tell you that I am {\it realistic\/} about is how little this note and your own efforts will help in setting the label game straight for QBism.  For 20 years (very literally now), I have pleaded with the community to understand my efforts at understanding quantum mechanics as being part of a realist program, i.e., as an attempt to say something about what the world is like, how it is put together, and what's the stuff of it.  But I have failed miserably at getting nearly anyone to understand this.  You are very likely to fail too, no matter how accurate you ultimately make your chart.  So many in our field---Matt Leifer, Tim Maudlin, David Albert, Avshalom Elitzur, Harvey Brown, Adrian Kent, Terry Rudolph, Travis Norsen, Sean Carroll, Paul Davies, Lee Smolin, Lucien Hardy, are a small sampling that come to mind---not to mention troves of science journalists, are simply unwilling to register the distinctions needed to have this discussion.  These people are so stuck in a pre-conceived worldview, they cannot see the slightest bit out of it, even when approached by a nonstandard one of their own (i.e., a realist of a different flag).

That said, the only person in your present citation list who I would \underline{not} call a realist is Asher Peres.  Asher, in fact, took pride in calling himself alternately a positivist and an instrumentalist.  Here are two instances where he labeled himself as such in print:  \myurl{http://arxiv.org/abs/quant-ph/9711003} and \myurl{http://arxiv.org/abs/quant-ph/0310010}.  (You may note that he labelled me with the same terms as well.  This was one of the key issues that made the writing of our joint article in {\sl Physics Today} \cite{Fuchs00a} so frustrating; every single sentence had to be a careful negotiation in language so that I could feel I wasn't selling my soul.)  Asher was fully happy in thinking that the task of physics was solely in making better predictions from sense data to sense data.

Here's the way, I view my own realism.  It can be taken straight from the playbook of Albert Einstein.  He wrote in the ``Reply to Criticisms'' section of the Schilpp volume \cite{Einstein49}:
\bq
A few more remarks of a general nature concerning concepts and [also] concerning the insinuation that a concept---for example that of the real---is something metaphysical (and therefore to be rejected).  A basic conceptual distinction, which is a necessary prerequisite of scientific and pre-scientific thinking, is the distinction between ``sense-impressions'' (and the recollection of such) on the one hand and mere ideas on the other.  There is no such thing as a conceptual definition of this distinction (aside from circular definitions, i.e., of such as make a hidden use of the object to be defined).  Nor can it be maintained that at the base of this distinction there is a type of evidence, such as underlies, for example, the distinction between red and blue.  Yet, one needs this distinction in order to be able to overcome solipsism.  Solution:  we shall make use of this distinction unconcerned with the reproach that, in doing so, we are guilty of the metaphysical ``original sin.''  We regard the distinction as a category which we use in order that we might the better find our way in the world of immediate sensations.  The ``sense'' and the justification of this distinction lies simply in this achievement.  But this is only a first step.  We represent the sense-impressions as conditioned by an ``objective'' and by a ``subjective'' factor.  For this conceptual distinction there also is no logical-philosophical justification.  But if we reject it, we cannot escape solipsism.  It is also the presupposition of every kind of physical thinking.  Here too, the only justification lies in its usefulness.  We are here concerned with ``categories'' or schemes of thought, the selection of which is, in principle, entirely open to us and whose qualification can only be judged by the degree to which its use contributes to making the totality of the contents of consciousness ``intelligible.''  The above mentioned ``objective factor'' is the totality of such concepts and conceptual relations as are thought of as independent of experience, viz., of perceptions.  So long as we move within the thus programmatically fixed sphere of thought we are thinking physically.  Insofar as physical thinking justifies itself, in the more than once indicated sense, by its ability to grasp experiences intellectually, we regard it as ``knowledge of the real.''

After what has been said, the ``real'' in physics is to be taken as a type of program, to which we are, however, not forced to cling a priori.  ...

The theoretical attitude here advocated is distinct from that of Kant only by the fact that we do not conceive of the ``categories'' as unalterable (conditioned by the nature of the understanding) but as (in the logical sense) free conventions.  They appear to be a priori only insofar as thinking without the positing of categories and of concepts in general would be as impossible as breathing in a vacuum.
\eq

If I were to modify the wording of this any at all, I would emphasize the word ``experience'' in place of Einstein's ``sensations'' and ``sense-impressions,'' and I would also probably give a more Darwinian-toned story about survival, etc., in place of Einstein's ``we might better find our way in the world.''  But these are minor things.  A little more serious would be that I {\it do\/} think we have a kind of direct evidence of ``the real.''  It is in the very notion of experience itself; I'll come back to this later.  What I want to emphasize at the moment is that I cannot see any way in which the program of QBism has ever contradicted what Einstein calls the program of ``the real.''

The only ``sin'' that QBism has ever committed is that it has held forthrightly and obstinately to the thought that the best understanding of quantum theory is obtained by recognizing that quantum states, quantum time-evolution maps, and the outcomes of quantum measurements all live within what Einstein calls the ``subjective factor.''  But then it is just a {\it non sequitur\/} for anyone to say, ``Well then everything in quantum theory, nay everything in physics, must live in the subjective factor.  Solipsism!  Non-realism!  Anti-realism!  Mysticism!  QBists don't believe in reality!  REALITY IS AN ILLUSION! (the headlines say).''  This is because, if any of these cads were to take a moment to think about it, they would recognize that there is more to quantum mechanics than just three isolated terms (states, evolution, and measurement)---there's the full-blown theory that glues these notions together in a very particular way, and in a way that would have never been discovered without empirical science.

It is from this glue that QBism draws the ``free conventions'' Einstein speaks of.  Take for example the Born Rule, expressed in the way R\"udiger and I find its content most revealing---for example, Eq.\ (8) in \myurl{http://arxiv.org/abs/1003.5209},
$$
Q(D_j)\; =\; (d+1)\sum_{i=1}^{d^2} P(H_i) P(D_j|H_i) - 1\;.
$$
We call this a normative rule:  It is a rule that any agent should strive for in his probability assignments with regard to a system he judges to have dimension $d$.  It doesn't set or fix the values of the probabilities---those depend upon the agent's prior personal experiences, his computational powers, and a good amount of guesswork---but the equation itself is something he should {\it strive\/} to satisfy with his gambling commitments (i.e., probability assignments).  In that sense, the Born Rule is much larger, less agent-specific than any probability assignments themselves.  It's a rule that any agent should pick up and use.  To that extent, it lives in what Einstein calls the ``objective factor''---it lives at the level of the impersonal.  And because of that, the Born Rule correlates with something that one might want to call ``real,'' as justified only (Einstein) by the way we might use it to ``better find our way in the world.''\footnote{It is worth noting that {\it in this aspect\/} at least, QBism bears a certain resemblance to structural realism.  See, for instance, \myurl{http://plato.stanford.edu/entries/structural-realism/}. Imagine our universe at a time (if there ever was one) when there were no agents about to use the laws of probability theory as an aid in their gambles---i.e., no such agents had yet arrived out of the Darwinian goo.  Were there any quantum states in the universe then?  A QBist would say NO\@. It's not a matter of the quantum state of the universe waiting until a qualified PhD student came along before having its first collapse, as John Bell joked, but that there simply weren't any quantum states.  Indeed, on earth there weren't any quantum states until 1926 when Erwin Schr\"odinger wrote the first one down.  The reason is simple: The universe is made of something else than quantum states.  But then, what of the Born Rule?  To this, in contrast, a QBist would say, ``Aha, now there's a sensible question.'' For the Born Rule is among the set of relations an agent should strive to attain in his larger mesh of probability assignments.  That normative rule is still lying about even when there are no agents to make use of it.  As Craig Callender once paraphrased it back to me in a conversation, it's the normative rule which is nature's whisper, not the specific terms within it.}

So, that's one direction in which QBism points to realism.  But there are other directions, and these have more to do with some of the things possibly nearer to your heart (and what you had wanted to use as a distinguishing factor between QBism and Wheeler-Zeilinger-Brukner).  Go back to the Eq.\ (8) advertised above.  An easy consequence of this normative rule is that only in very limited circumstances can one make sharp (i.e., delta-function) probability assignments.  Another way to say this is that quantum theory itself advises us that we should not think we can always have certainties.  This, I believe, is what you're wanting to call ``irreducible randomness,'' but it's nuanced in ways that I think Brukner and Zeilinger might not take note of.  On the one hand, our Eq.\ (8) immediately gives something like what Rob Spekkens calls an ``epistemic restriction'' and Zeilinger calls an ``information principle,'' but one has to remember that QBism is concerned {\it not\/} with ``epistemic probabilities'' (i.e., as only statements of ignorance or how much is known) or with some kind of spiritual substance in the way that Zeilinger often speaks (for instance when he says ``the world is {\it made\/} of information''), but with probabilities as guides to action---de Finetti and Ramsey's notion of probability.  In this way, QBist probabilities have no direct connection with what nature ``must do,'' not even when $p=1$.  And that is a hugely important distinction between QBism and any other view of quantum theory.  David Mermin  stated this point very clearly in \myurl{http://arxiv.org/abs/1409.2454}:
\bq
A very important difference of QBism, not only from Copenhagen, but from virtually all other ways of looking at science, is the meaning of probability 1 (or 0).  In Copenhagen quantum mechanics, an outcome that has probability 1 is enforced by an objective mechanism.  This was most succinctly put by Einstein, Podolsky and Rosen, though they were, notoriously, no fans of Copenhagen.  Probability-1 judgments, they held, were backed up by ``elements of physical reality''.

Bohr held that the mistake of EPR lay in an ``essential ambiguity'' in their phrase ``without in any way disturbing''.  For a QBist, their mistake is much simpler than that:  probability-1 assignments, like more general probability-$p$ assignments are personal expressions of a willingness to place or accept bets, constrained only by the requirement that they should not lead to certain loss in any single event.  It is wrong to assert that probability assignments must be backed up by objective facts on the ground, even when $p=1$.  An expectation is assigned probability 1 if it is held as strongly as possible. Probability-1 measures the intensity of a belief: supreme confidence.  It does not imply the existence of a deterministic mechanism.

We are all used to the fact that with the advent of quantum mechanics, determinism disappeared from physics.    Does it make sense for us to qualify this in a footnote: ``Except when quantum mechanics assigns probability 1 to an outcome''?
\eq
This is what R\"udiger meant in his note to you when he said, ``QBism thus takes the idea of irreducible randomness much further than Zeilinger or Brukner.''\footnote{R\"udiger Schack has furthermore found a very fine way of putting this point in some recent talks delivered in Stellenbosch and Siegen. First he quotes Einstein, in a letter to F.~Reiche and his wife, dated 15 August 1942.  Einstein wrote, ``I still do not believe that the Lord God plays dice. If he had wanted to do this, then he would have done it quite thoroughly and not stopped with a plan for his gambling: In for a penny, in for a pound. Then we wouldn't have to search for laws at all.'' Then, on the next slide R\"udiger writes, ``The usual reading: Einstein advocates deterministic laws.  QBist reading: There are indeed no laws. {\it God has done it thoroughly}. There are no laws of nature, not even stochastic ones. The world does not evolve according to a mechanism.  What God has provided, on the other hand, is tools for agents to navigate the world, to survive in the world.''}  The way I might put it is that quantum theory advises us to recognize that no matter how we slice up nature, we will never find any pieces of it beholden to {\it laws\/} in the usual physics sense of the term.  John Wheeler toyed with the idea that ``the only law of nature is that there is no law,'' and quantum theory on a QBist reading supports this.

So, what backs up this position?  The answer is:  The general structure of quantum theory---i.e., the part correlative with ``the real'' (in the Einstein sense of a ``free convention'')---in particular the part of it you have explored so thoroughly, the Kochen-Specker and Bell theorems in all their great variety.  Thus, part of your chart is definitely wrong in how it attempts to draw a distinction between QBism and Wheeler-Zeilinger-Brukner.  QBism agrees with Wheeler that quantum measurements gives rise to new creation within the universe---he likened these creations to the big bang itself, see our \myurl{http://arxiv.org/abs/1412.4209}---but QBism maybe even goes further than Wheeler would have, I don't know, in that it holds fast to this statement even when one has a $p=1$ prediction beforehand.  Now, going that far---i.e., asserting no pre-existent properties even when $p=1$---is indeed a point of distinction between Brukner, Zeilinger, and Kofler and QBism.  You can find instances of this statement in several places in their writings (what the philosophers call the ``eigenstate-eigenvalue link'') \cite{Brukner01,Kofler10}.  Another place where your table is off the mark connects to what I alluded to above when I said I have a bigger disagreement with the Einstein quote.  You cite a 2005 Zeilinger paper \cite{Zeilinger05a} for saying, ``this randomness of the individual event is the strongest indication we have of a reality `out there' existing independently of us,'' but you could just as well have cited my 2002 paper \myurl{http://arxiv.org/abs/quant-ph/0204146}:

\bq
I would say all our evidence for the reality of the world comes from without us, i.e., not from within us.  We do not hold evidence for an independent world by holding some kind of transcendental knowledge.  Nor do we hold it from the practical and technological successes of our past and present conceptions of the world's essence.  It is just the opposite.  We believe in a world external to ourselves precisely because we find ourselves getting unpredictable kicks (from the world) all the time.  If we could predict everything to the final T as Laplace had wanted us to, it seems to me, we might as well be living a dream.

To maybe put it in an overly poetic and not completely accurate way, the reality of the world is not in what we capture with our theories, but rather in all the stuff we don't.  To make this concrete, take quantum mechanics and consider setting up all the equipment necessary to prepare a system in a state $\Pi$ and to measure some noncommuting observable $H$.  (In a sense, all that equipment is just an extension of ourselves and not so very different in character from a prosthetic hand.)  Which eigenstate of $H$ we will end up getting as our outcome, we cannot say.  We can draw up some subjective probabilities for the occurrence of the various possibilities, but that's as far as we can go.  (Or at least that's what quantum mechanics tells us.)  Thus, I would say, in such a quantum measurement we touch the reality of the world in the most essential of ways.
\eq

However, I don't like using the term ``intrinsic randomness'' for all this, and you detected as much in my interview with Schlosshauer \cite{Fuchs11a}.  That's because ``intrinsic randomness'' strikes me as such a lifeless phrase in comparison to what I think is the deeper issue---namely, that all the components/pieces/slices of the world have a genuine autonomy unto themselves.  And when those pieces get together, they go on to create even more autonomous stuff.  This way of speaking tries to capture that there's a creativity or novelty in the world in a way that the phrase ``intrinsic randomness'' leaves limp.  Here's how I put it in an introduction to a lecture I gave at Caltech in 2004 (reprinted in
\myurl{http://arxiv.org/abs/1405.2390}):

\bq
A lecturer faces a dilemma when teaching a course at a farsighted summer school like this one.  This is because, when it comes to research, there is often a fine line between what one thinks and what is demonstrable fact.  More than that, conveying to the students what one thinks---in other words, one's hopes, one's desires, the potentest of one's premises---can be just as empowering to the students' research lives (even if the ideas are not quite right) as the bare tabulation of any amount of demonstrable fact.  So I want to use one percent of this lecture to tell you what I think---the potentest of all my premises---and use the remaining ninety-nine to tell you about the mathematical structure from which that premise arises.

I think the greatest lesson quantum theory holds for us is that when two pieces of the world come together, they give birth.  [Bring two fists together and then open them to imply an explosion.]  They give birth to FACTS in a way not so unlike the romantic notion of parenthood:  that a child is more than the sum total of her parents, an entity unto herself with untold potential for reshaping the world.  Add a new piece to a puzzle---not to its beginning or end or edges, but somewhere deep in its middle---and all the extant pieces must be rejiggled or recut to make a new, but different, whole.\footnote{It is a bit of a stretch, but I have found a {\it wildly-speculative\/} idea in some recently unearthed notes from a 1974 notebook of John Wheeler's which is mildly evocative of the metaphor just given.  See \myurl{https://jawarchive.files.wordpress.com/2012/03/twa-1974.pdf} and \myurl{https://jawarchive.files.wordpress.com/2012/03/tarski.pdf}, typed transcripts of which may be found in the Appendix. Despite the dubious connection to anything firmly a part of QBism, I report Wheeler's idea because it seems to me that it conveys some imaginative sense of how the notion of ``birth'' described here carries a very different flavor from the ``intrinsic randomness'' that Ad\'an and others seem to be talking about.

Imagine along with Wheeler that the universe can somehow be identified with a formal mathematical system, with the universe's life somehow captured by all the decidable propositions within the system.  Wheeler's ``crazy'' idea seems to be this.  Every time an act of observer-participancy occurs (every time a quantum measurement occurs), one of the {\it un}decidable propositions consistent with the system is upgraded to the status of a new axiom with truth value either TRUE or FALSE\@. In this way, the life of the universe as a whole takes on a deeply new character with the outcome of each quantum measurement. The ``intrinsic randomness'' dictated by quantum theory is not so much like the flicker of a firefly in the fabric of night, but a rearrangement of the whole meaning of the universe.}  That is the great lesson.

But quantum mechanics is only a glimpse into this profound feature of nature; it is only a part of the story.  For its focus is exclusively upon a very special case of this phenomenon:  The case where one piece of the world is a highly-developed decision-making agent---an experimentalist---and the other piece is some fraction of the world that captures his attention or interest.

When an experimentalist reaches out and touches a quantum system---the process usually called quantum `measurement'---that process gives rise to a birth.  It gives rise to a little act of creation.  And it is how those births or acts of creation impact the agent's expectations for other such births that is the subject matter of quantum theory.  That is to say, quantum theory is a calculus for aiding us in our decisions and adjusting our expectations in a QUANTUM WORLD.  Ultimately, as physicists, it is the quantum world for which we would like to say as much as we can, but that is not our starting point.  Quantum theory rests at a level higher than that.

To put it starkly, quantum theory is just the start of our adventure.  The quantum world is still ahead of us.  So let us learn about quantum theory.
\eq

Afterward, again in \myurl{http://arxiv.org/abs/1405.2390}, you'll find a lot of discussion about how the word ``FACTS'' is not quite right for these little moments of creation (that word itself is a bit too lifeless for what I am hoping to convey)---look up the words QBoom and QBlooey:

\vspace*{-15pt}
\bq
\noindent
\begin{itemize}
\item
{\bf QBoom} -- cf.~{\it kaboom}; the sought-for deanthropocentrized distillate of quantum measurement that QBism imagines powering the world.  William James called it ``pure experience,'' where ``new being comes in local spots and patches which add themselves or stay away at random, independently of the rest.''  John Wheeler asked, ``Is the entirety of existence, rather than built on particles or fields or multidimensional geometry, built on billions upon billions of elementary quantum phenomena, those elementary acts of `observer-participancy'? \ldots\ Is what took place at the big bang the consequence of billions upon billions of these elementary `acts of observer-participancy'?''  In place of a Big Bang, the QBist wonders whether it might not be myriads and myriads of little QBooms!

\item
{\bf QBlooey} -- cf.~{\it kablooey}; a QBist slur on the usual conception of the Big Bang, where the universe had its one and only creative moment at the very beginning.
\end{itemize}
\eq
but I hope this is enough to give you at least a hint of why I don't like the flavor of the term ``intrinsic randomness'' even if there is a technical sense in which it is correct to QBism.

I thought a little about why you might have thought QBism was noncommittal on this idea of ``intrinsic randomness.''  Perhaps it was because of this slide which I presented in Buenos Aires:

\bq
\begin{center}
{\bf No Commitment to Ontology Here}\medskip
\end{center}

Most of the time one sees Bayesian probabilities characterized as measures of ignorance or imperfect knowledge.  But that description carries with it a metaphysical commitment that is not necessary for the personalist Bayesian.

Imperfect knowledge?  It sounds like something that, at least in imagination, could be perfected, making all probabilities zero or one---one uses probabilities only because one does not know the true pre-existing state of affairs.

All that matters is that there is {\it uncertainty\/} for whatever reason. There might be uncertainty because there is ignorance of a true state of affairs, but there might be uncertainty because the world itself does not yet know what it will give---i.e., there is an objective indeterminism.
\eq

If so, then maybe this will help dispel the confusion:  You should understand that the slide was only meant to make a statement about the subjective Bayesian notion of probability.  The notion itself is noncommittal.  However, subjective Bayesian probability {\it combined\/} with the Born Rule and Kochen-Specker considerations, etc., is very committal.  In fact, I drew the words for that slide from Footnote 14 of \myurl{http://arxiv.org/abs/1003.5209}, but what I left out was the footnote's very ending sentence:  ``As will be argued in later sections, QBism finds its happiest spot in an unflinching combination of `subjective probability' with `objective indeterminism'.''  See also this abstract from another talk where I point out how QBism's metaphysics may not have been amenable to de Finetti's own:

\bq
\begin{center}
{\bf Something on QBism}\medskip
\end{center}

The term QBism, invented in 2009, initially stood for Quantum Bayesianism, a view of quantum theory a few of us had been developing since 1993.  Eventually, however, I. J. Good's warning that there are 46,656 varieties of Bayesianism came to bite us, with some Bayesians feeling their good name had been hijacked.  David Mermin suggested that the B in QBism should more accurately stand for ``Bruno,'' as in Bruno de Finetti, so that we would at least get the variety of (subjective) Bayesianism right.  The trouble is QBism incorporates a kind of metaphysics that even Bruno de Finetti might have rejected!  So, trying to be as true to our story as possible, we momentarily toyed with the idea of associating the B with what the early 20th-century United States Supreme Court Justice Oliver Wendell Holmes Jr.\ called ``bettabilitarianism.''  It is the idea that the world is loose at the joints, that indeterminism plays a real role in the world.  In the face of such a world, what is an active agent to do but {\it participate\/} in the uncertainty that is all around him?  As Louis Menand put it, ``We cannot know what consequences the universe will attach to our choices, but we can bet on them, and we do it every day.''  This is what QBism says quantum theory is about:  How to best place bets on the consequences of our actions in this quantum world.  But what an ugly, ugly word, {\it bettabilitarianism\/}!  Therefore, maybe one should just think of the B as standing for no word in particular, but a deep idea instead:  That the world is so wired that our actions as active agents actually {\it matter}.  Our actions and their consequences are not eliminable epiphenomena.  In this talk, I will describe QBism as it presently stands and give some indication of the many things that remain to be developed.
\eq

So, I write all these words to say, {\it PLEASE\/} don't call us ``non-realist'' in print.  We have endured enough hardship because of that damned label.  That label probably won't ever leave us (my first realism expressed above), but I would like to think that you, our friend, never contributed any of your own to the trouble.

Then, you say, ``Well then, what instead should I call the category in which QBism sits?  If not non-realism, what?''  I have an answer.  But before I propose it, I want to say something about one more {\it realism\/} of QBism.  By lifting the quantum formalism from being directly descriptive of the world external to the agent and instating it instead as a normative prescription for aiding in his survival, at the same time as asserting that agents are physical systems like any others, QBism breaks into a territory the vast majority of those declaring they have a scientific worldview would be loath to enter.  And that is that the agents (observers) matter as much as electrons and atoms in the construction of the actual world---the agents using quantum theory are not incidental to it.  Johannes Kofler said recently to R\"udiger Schack, ``I am convinced we are biological machines.''  I told R\"udiger his reply should have been, ``The lesson of QBism is that there are {\it no\/} machines, period.''  In the terms that Danny Greenberger laid out in \myurl{http://www.iqoqi-vienna.at/can-a-computer-ever-become-conscious/}, everything in the universe has the potential to ``push the big red button.''  And nothing should be forgotten in that regard, from the lowliest electron to the very user of quantum theory.  I've always admired the way my student John DeBrota put the key point in his graduate application:  ``From [QBism] I have learned that the universe can be moved---that it is not a masterfully crafted mechanical automaton, but instead an unfinished book, brimming with creation and possibility, of which every agent is an author.''  If that too is a reality of QBism, then it should be recognized in its categorization.

It's the reality that the observer, the agent is an active and non-negligible {\it pariticipator\/} in the universe as John Wheeler emphasized so many times.  Certainly, recall Wheeler's game of 20 questions, surprise version (taken from ``Frontiers of Time,'' 1979) \cite{Wheeler79b}:

\bq
The Universe can't be Laplacean.  It may be higgledy-piggledy.  But have hope.  Surely someday we will see the necessity of the quantum in its construction.  Would you like a little story along this line?

Of course!  About what?

About the game of twenty questions.  You recall how it goes---one of the after-dinner party sent out of the living room, the others agreeing on a word, the one fated to be a questioner returning and starting his questions.  ``Is it a living object?''  ``No.''  ``Is it here on earth?''  ``Yes.''  So the questions go from respondent to respondent around the room until at length the word emerges: victory if in twenty tries or less; otherwise, defeat.

Then comes the moment when we are fourth to be sent from the room.  We are locked out unbelievably long.  On finally being readmitted, we find a smile on everyone's face, sign of a joke or a plot.  We innocently start our questions.  At first the answers come quickly.  Then each question begins to take longer in the answering---strange, when the answer itself is only a simple ``yes'' or ``no.''  At length, feeling hot on the trail, we ask, ``Is the word `cloud'?''  ``Yes,'' comes the reply, and everyone bursts out laughing.  When we were out of the room, they explain, they had agreed not to agree in advance on any word at all.  Each one around the circle could respond ``yes'' or ``no'' as he pleased to whatever question we put to him.  But however he replied he had to have a word in mind compatible with his own reply---and with all the replies that went before.  No wonder some of those decisions between ``yes'' and ``no'' proved so hard!

And the point of your story?

Compare the game in its two versions with physics in its two formulations, classical and quantum.  First, we thought the word already existed ``out there'' as physics once thought that the position and momentum of the electron existed ``out there,'' independent of any act of observation.  Second, in actuality the information about the word was brought into being step by step through the questions we raised, as the information about the electron is brought into being, step by step, by the experiments that the observer chooses to make. Third, if we had chosen to ask different questions we would have ended up with a different word---as the experimenter would have ended up with a different story for the doings of the electron if he had measured different quantities or the same quantities in a different order.  Fourth, whatever power we had in bringing the particular word ``cloud'' into being was partial only.  A major part of the selection---unknowing selection---lay in the ``yes'' or ``no'' replies of the colleagues around the room.  Similarly, the experimenter has some substantial influence on what will happen to the electron by the choice of experiments he will do on it; but he knows there is much impredictability about what any given one of his measurements will disclose.  Fifth, there was a ``rule of the game'' that required of every participator that his choice of yes or no should be compatible with some word.  Similarly, there is a consistency about the observations made in physics.  One person must be able to tell another in plain language what he finds and the second person must be able to verify the observation.
\eq

In honor of that, and because I do think it captures a big part of the idea we are all expressing, I would suggest calling our category ``Participatory Realism.''  I would say everyone you presently categorize as non-realist (with the exception of Asher) partakes to some extent in a kind of participatory realism.  Wheeler has more of it than Zeilinger, and Zeilinger may have more of it than Bohr, and Bohr more than Rovelli, but they all have a bit of it.  QBism carries it to its logically enforced extreme.

Finally, if QBism and the others in the right-side table are categorized as ``participatory realism,'' the ones on the left side can't be called simply realist.  For they are a very specialized species of the genus:  All they really want is their hidden variable, somewhere, somehow (Spekkens too is certainly guilty of this even though he would try to deny it).  And they want physicists and humankind more generally to be relegated to being inessential epiphenomena in the universe.  They want the physicist to really have nothing to with Physics with a capital P.  Thus, here are some suggestions:
\bv
{\it dumb realism\/} \ \ (``lacking intelligence or good judgment; stupid; dull-witted'')\\
{\it static realism}\\
{\it stillborn realism}\\
{\it base realism\/} \ \ (``morally low; without estimable qualities; dishonorable; cowardly'')
\ev
But whatever you do, {\it PLEASE\/} call QBism ``{\it any damned thing but}'' non-realist!  (See this video of Johnny Cash singing ``A Boy Named Sue,'' starting at the 2:53 mark of \myurl{https://www.youtube.com/watch?v=WOHPuY88Ry4}.)

\section{``Slicing the Euglena's Tail,'' to Ad\'an Cabello, 28 July 2015}

And one further note to address this remark you made to R\"udiger in your last note:
\bac
This may be the point!  For QBism, quantum theory is ``a `user's manual' any
agent can pick up.''  End of the story.  While Zeilinger is trying to get insight
about how nature works from that manual.  Am I right?
\eac

You are wrong.  I told the story below in my talk in Bueno Aires, but it probably went by too fast for you to catch.  See the missing paragraphs below from my interview with Amanda Gefter: \myurl{https://www.quantamagazine.org/20150604-quantum-bayesianism-qbism/}. They address your contention head on.

\bq
\noindent {\bf Amanda:}  QBism says that quantum mechanics is a ``single-user theory.''\medskip

\noindent {\bf Chris:}  Any of us can use quantum theory, but only for ourselves. There's a little single-celled thing called a Euglena that has a tail coming off of it. The tail arose from evolutionary pressures, so that the Euglena can move from environments where there are depleted nutrients to environments where there's an abundance of nutrients. It's a tool. Quantum mechanics is like the Euglena's tail. It's something we evolved in 1925 and since it's been shown to be such a good tool, we keep using it and we pass it on to our children. The tail is a single-user tail. But we can look at the tail and ask things like, what might we learn about the environment by studying its structure? We might notice the tail is not completely circular and that might tell us something about the viscosity of the medium it's traveling through. We might look at the ratio of the length of it to the width of it in various places and that might tell us about features of the environment. So quantum mechanics is a single-user theory, but by dissecting it, you can learn something about the world that all of us are immersed in.\medskip

\noindent {\bf Amanda:}  So eventually objectivity comes in?\medskip

\noindent {\bf Chris:}  I hope it does. Ultimately I view QBism as a quest to point to something in the world and say, that's intrinsic to the world.  But I don't have a conclusive answer yet. Let's take the point of view that quantum mechanics is a user's manual. A user's manual {\it for me}. A philosopher will quickly say, well that's just instrumentalism. ``Instrumentalism'' is always prefaced by a ``just.'' But that's jumping too quickly to a conclusion. Because you can always ask---you should always ask---what is it about the world that compels me to adopt this instrument rather than that instrument? A quantum state is a user's manual of probabilities. But how does it determine the probabilities? Well there's a little mathematical formula called the Born rule. And then you should ask, why that formula? Couldn't it have been a different formula? Yes, it might have been different. The fact that we adopt this formula rather than some other formula is telling us something about the character of the world as it is, independent of us. If we can answer the question ``Why the Born Rule?''\ or John Wheeler's question ``Why the quantum?''\ then we'll be making a statement about how the world is, one that's not ``just'' instrumentalism.
\eq

\section{``Denouement,'' to Johannes Kofler, 6 October 2014}

Here is the result that came out of your kindness of lending me your Schilpp volume and letting me subject its spine to the copier.

Let me just tell you a bit about how this little quest fits into my bigger agenda (and is something I view as a mandate for QBism).  It's that I take absolutely seriously John Wheeler's ``idea for an idea'' (as he would say) that the ``elementary quantum phenomenon'' might be taken as the ultimate building block of reality.  I also take absolutely seriously his idea that within every ``phenomenon'' is an instance of creation, not unlike what one usually exclusively associates with the Big Bang.  This caused John to speculate that ``perhaps the Big Bang is here all around us.''  (This idea that {\it every\/} quantum measurement results in an instance of creation is also connected to the QBist rejection of the EPR reality criterion.  Even outcomes following probability-1 assignments are for us are instances of creation.  We could not have that if probability-1 in fact meant pre-existence.  But this is an aside, connected more to the note I will send you following this one.)

So getting Bohr's idea of ``phenomenon'' straight to myself is quite important from my perspective.  I think much of what Bohr says was flawed, but that doesn't mean I still don't have several things to learn from him.  What he calls ``phenomenon'' resembles in many ways what a set of philosophers early last century (William James, John Dewey, Ralph Barton Perry, and some others) called ``pure experience'' \ldots\ though they explored their idea from all angles and in great detail, rather than spewing the same few phrases over and over like Bohr did.  Importantly James saw his notion of ``pure experience'' as the ultimate building block of reality---i.e., like Wheeler of Bohr's ``phenomenon.''  The reason Bohr still interests me, however, even though his story is hardly written at all compared to the great corpus of James and Dewey, is that he at least makes a connection to the quantum, whereas they could not.  In fact, as best I can tell, when Bohr continually talks about the indivisibility of the quantum, he is using that as a synonym for what he calls in other contexts ``the quantum postulate.''  So he saw the ``phenomena'' as the very reason for being for quantum mechanics.

The greatest and most difficult task will be---after developing QBism sufficiently well to get a clear view of all that is around it---to return to what Bohr tried to jump to prematurely with his ``classically described agencies of observation'' and {\it disambiguate\/} the ``pure experience'' notion from the agents gambling upon them.  Only then can one imagine ``pure experience'' or ``phenomenon'' as an ``ultimate building block for all reality.''

\bigskip\bigskip
\noindent {\Large \bf Acknowledgements}\bigskip

I thank Ted H\"ansch and Ignacio Cirac for affording me the leisure to think on these things in the quiet of little Garching some time before writing them down.  I thank Blake Stacey for advice on this manuscript.\medskip

\bigskip\bigskip
\noindent {\Large \bf Appendix: Transcription from John Wheeler's Notebook}\bigskip

\noindent [Text taken from photographs of notebook posted at \myurl{https://jawarchive.wordpress.com/}.]\bigskip\medskip

\noindent 4--6 February 1974 \underline{draft notes} for discussion with Dana Scott; also with Simon Kochen, Charles Patton and Roger Penrose.

\begin{center}
ADD ``PARTICIPANT'' TO ``UNDECIDABLE PROPOSITIONS'' TO ARRIVE AT PHYSICS \medskip
\\
John A. Wheeler
\end{center}

\begin{center}
\underline{Brief}
\end{center}

We consider the quantum principle.  Of all of the well-analyzed features of physics there is none more central, and none of an origin more mysterious.  This note identifies its key idea as \underline{participation}; and its point of origin, as the ``undecidable propositions'' of mathematical logic.  On this view physics is not machinery. Logic is not oil occasionally applied to that machinery. Instead everything, physics included, derives from two parents, and is nothing but cathode-tube image of the interplay between them. One is the ``participant.'' The other is the complex of undecidable propositions of mathematical logic. The ``participator'' assigns true-false values to appropriate ones among these propositions at his own free will. As he does so, the corresponding world unrolls on his screen. No participator, no world!

\begin{center}
\underline{Comments}
\end{center}

\noindent 1. \underline{The quantum principle and physics.} \medskip

The quantum principle is taken here to be the pervasive unifying element of all of physics. It would be a favorable sign to find the quantum principle derivable from mathematical logic along the foregoing lines; and to find the opposite would be a decisive blow against these views.\medskip

\noindent 2. \underline{Start with what formal system?} \medskip

\noindent Take a formal system. Enlarge it to a new formal system, and that again to a new formal system, and so on, by resolving undecidable propositions (``act of participation'').  Will the system become so complex that it can and must be treated by statistical means?  Will such a treatment make it irrelevant, or largely irrelevant, with what particular formal system one started?\medskip

\noindent 3. \underline{Lack of commutativity} \medskip

\noindent After the system has grown to a certain level of complexity, one can imagine a difference in the subsequent development, according as decisions about appropriate ``undecidable propositions'' are taken in the order AB or the order BA.  One might focus on this point in trying to locate something like the quantum principle as already contained in mathematical logic.\medskip

\noindent 4. \underline{``Reality''} \medskip

\noindent The propositions are not propositions about anything. They are the abstract building blocks, or ``pregeometry,'' out of which ``reality'' is conceived as being built.\medskip\bigskip

\noindent Mon 25 Feb '74 \ \underline{PATTON, FOLLOW-UP OF SCOTT DISCUSSION} \bigskip

Can we find ``Tr'' in our theory and A in our theory such that for all provable statements it comes out OK and yet also to unprov.\ statements assign truth values[?]  Tarski says can't do; have to have a bigger theory; have to have someone on outside \underline{imposing} what's true \& what's false.  Truth is thus a ``meta'' concept.  ``Participator'' here!  Logic can't live by itself.  No wonder Boolean system won't fly.


\begin{thebibliography}{100}

\bibitem{Fuchs10a}
C.~A. Fuchs, ``QBism, the Perimeter of Quantum Bayesianism,'' {\tt arXiv:1003.5209 [quant- ph]} (2010).

\bibitem{Fuchs02a}
C.~A. Fuchs, ``Quantum Mechanics as Quantum Information (and only a little more),'' in {\sl Quantum Theory:\ Reconsideration of
Foundations}, edited by A.~Khrennikov (V\"axj\"o University Press, V\"axj\"o, Sweden, 2002), pp.~463--543; {\tt arXiv:quant-ph/0205039}.

\bibitem{Fuchs02b}
C.~A. Fuchs, ``The Anti-V\"axj\"o Interpretation of Quantum
Mechanics,'' in {\sl Quantum Theory:\ Reconsideration of Foundations},
edited by A.~Khrennikov (V\"axj\"o University Press, V\"axj\"o, Sweden, 2002), pp.~99--116; {\tt arXiv:quant-ph/0204146}.

\bibitem{Fuchs07a}
C.~A. Fuchs, ``Delirium Quantum: Or, where I will take quantum mechanics if it will let me,'' in {\sl Foundations of Probability and Physics -- 4}, edited by G.~Adenier, C.~A. Fuchs, and A.~Yu.\ Khrennikov, AIP Conference Proceedings Vol.~889, (American Institute of Physics, Melville, NY, 2007), pp.~438--462; {\tt arXiv:0906.1968 [quant-ph]}.

\bibitem{Fuchs11a}
C. A. Fuchs, ``Interview with a Quantum Bayesian,'' in {\sl Elegance and Enigma:\ The Quantum Interviews}, edited by M.~Schlosshauer (Springer, Berlin, Frontiers Collection, 2011); {\tt arXiv:1207.2141 [quant-ph]}.

\bibitem{Caves02a}
C.~M. Caves, C.~A. Fuchs and R.~Schack, ``Quantum Probabilities as Bayesian Probabilities,'' Phys.\ Rev.\ A {\bf 65}, 022305 (2002).

\bibitem{Caves02b}
C.~M. Caves, C.~A. Fuchs and R.~Schack, ``Unknown Quantum States:\ The Quantum de Finetti Representation,'' J. Math.\ Phys.\ {\bf 43}, 4537--4559 (2002).

\bibitem{Caves02c}
C.~M. Caves, C.~A. Fuchs, and R.~Schack, ``Conditions for Compatibility of Quantum-State Assignments,'' Phys.\ Rev.\ A {\bf 66}, 062111 (2002).

\bibitem{Caves07}
C.~M. Caves, C.~A. Fuchs, and R.~Schack, ``Subjective Probability and Quantum Certainty,'' Stud.\ Hist.\ Phil.\ Mod.\ Phys.\ {\bf 38}, 255--274 (2007).

\bibitem{Fuchs00a}
C.~A. Fuchs and A. Peres, ``Quantum Theory Needs No `Interpretation','' Phys.\ Tod.\ {\bf 53}(3), 70--71 (2000).

\bibitem{Fuchs00b}
C.~A. Fuchs and A. Peres, ``Quantum Theory -- Interpretation, Formulation, Inspiration:\ Fuchs and Peres Reply,'' Phys. Today {\bf 53}(9), 14, 90 (2000).

\bibitem{Appleby05a}
D.~M. Appleby, ``The Bell-Kochen-Specker Theorem,'' Stud.\ Hist.\ Phil.\ Mod.\ Phys.\ {\bf 36}, 1--28 (2005).

\bibitem{Appleby05b}
D.~M. Appleby, ``Facts, Values and Quanta,'' Found.\ Phys.\ {\bf 35}, 627--668 (2005).

\bibitem{Appleby05c}
D.~M. Appleby, ``Probabilities Are Single-Case, or Nothing,'' Opt.\ Spectr.\ {\bf 99}, 447--456 (2005).

\bibitem{Appleby06a}
D.~M. Appleby, ``Concerning Dice and Divinity,'' in {\sl Foundations of Probability and Physics -- 4}, edited by G.~Adenier, C.~A. Fuchs, and A.~Yu.\ Khrennikov, AIP Conference Proceedings Vol.~889, (American Institute of Physics, Melville, NY, 2007), pp.~30--39; {\tt arXiv:quant-ph/0611261}.

\bibitem{Appleby13a}
D.~M. Appleby, ``Mind and Matter: A Critique of Cartesian Thinking,'' in {\sl The Pauli-Jung Conjecture and Its Impact Today}, edited by H.~Atmanspacher and C.~A. Fuchs, (Imprint Academic, Exeter, UK, 2014), pp.~7--36; {\tt arXiv:1305.7381v1} (2013).

\bibitem{Appleby15a}
D.~M. Appleby, C. A. Fuchs, B. C. Stacey, and H. Zhu, ``The Qplex: A Novel Arena for Reconstructing Quantum Theory,'' forthcoming (2016).

\bibitem{Appleby09a}
D.~M. Appleby, {\AA}.~Ericsson, and C.~A. Fuchs, ``Properties of QBist State Spaces,'' Found.\ Phys.\ {\bf 41}, 564--579 (2010).

\bibitem{Stacey14a}
B. C. Stacey, ``SIC-POVMs and Compatibility among Quantum States,'' {\tt arXiv:1404.3774 [quant-ph]} (2014).

\bibitem{Stacey14b}
B. C. Stacey, ``Von Neumann Was Not a Quantum Bayesian,'' {\tt arXiv:1412.2409 [physics. hist-ph]} (2014).

\bibitem{Stacey15a}
B. C. Stacey, {\sl Multiscale Structure in Eco-Evolutionary Dynamics}, PhD thesis, Brandeis University, 2015, Chapters 5 and 10; {\tt arXiv:1509.02958 [q-bio.PE]}.

\bibitem{vonBaeyer16a}
H. C. von Baeyer, {\sl QBism:\ The Future of Quantum Physics}, (Harvard University Press, Cambridge, MA, 2016).

\bibitem{Mermin09a}
N. D. Mermin, ``What's Bad About This Habit,'' Phys.\ Today {\bf 62}(5), 8--9 (2009); reprinted in N.~D. Mermin, {\sl Why Quark Rhymes with Pork and other Scientific Diversions}, (Cambridge University Press, Cambridge, UK, 2016), Chapter 30.

\bibitem{Mermin09b}
N. D. Mermin, ``Mermin Habitually Answers Opinions, Real and Abstract:\ Mermin Replies,'' Phys.\ Today {\bf 62}(9), 14--15 (2009).

\bibitem{Mermin12a}
N. D. Mermin, ``Quantum Mechanics:\ Fixing the Shifty Split,'' Phys.\ Today {\bf 65}(7), 8--10 (2012); reprinted in N.~D. Mermin, {\sl Why Quark Rhymes with Pork and other Scientific Diversions}, (Cambridge University Press, Cambridge, UK, 2016), Chapter 31.

\bibitem{Mermin12b}
N. D. Mermin, ``Measured Responses to Quantum Bayesianism:\ Mermin Replies,'' {\bf 65}(12), 12--15 (2012).

\bibitem{Mermin13a}
N. D. Mermin, ``Annotated Interview with a QBist in the Making,'' {\tt arXiv:1301.6551 [quant-ph]} (2013).

\bibitem{Mermin13b}
N. D. Mermin, ``Impressionism, Realism, and the Aging of Ashcroft and Mermin:\ Reply to Men\'endez,'' Phys.\ Today {\bf 66}(7), 8 (2013).

\bibitem{Mermin14a}
N. D. Mermin, ``What I Think about Now,'' Phys.\ Today {\bf 67}(3), 8--9 (2014); reprinted in N.~D. Mermin, {\sl Why Quark Rhymes with Pork and other Scientific Diversions}, (Cambridge University Press, Cambridge, UK, 2016), Chapter 32.

\bibitem{Mermin14b}
N. D. Mermin, ``Classical and Quantum Framing of the Now:\ Mermin Replies,'' Phys.\ Today {\bf 67}(9), 8--9 (2014).

\bibitem{Mermin14c}
N. D. Mermin, ``QBism Puts the Scientist Back into Science,'' Nature {\bf 507}, 421--423 (2014).

\bibitem{Mermin14d}
N. D. Mermin, ``QBism in the New Scientist,'' {\tt arXiv:1406.1573 [quant-ph]} (2014).

\bibitem{Fuchs14a}
C.~A. Fuchs, N. D. Mermin, and R.~Schack, ``An Introduction to QBism with an Application to the Locality of Quantum Mechanics,'' Am.\ J. Phys.\ {\bf 82}, 749--754 (2014).

\bibitem{Mermin14f}
N. D. Mermin, ``Why QBism is Not the Copenhagen Interpretation and What John Bell Might Have Thought of It,'' {\tt arXiv:1409.2454 [quant-ph]} (2014); reprinted in N.~D. Mermin, {\sl Why Quark Rhymes with Pork and other Scientific Diversions}, (Cambridge University Press, Cambridge, UK, 2016), Chapter 33.

\bibitem{Fuchs15a}
C.~A. Fuchs, N. D. Mermin, and R.~Schack, ``Reading QBism:\ Reply to Nauenberg,'' Am.\ J. Phys.\ {\bf 83}, 198 (2015).

\bibitem{Schack01a}
R. Schack, T. A. Brun, and C. M. Caves, ``Quantum Bayes Rule,'' Phys.\ Rev.\ A {\bf 64}, 014305 (2001).

\bibitem{Fuchs04a}
C.~A. Fuchs, R.~Schack, and P.~F. Scudo, ``A de Finetti
Representation Theorem for Quantum Process Tomography,'' Phys.\
Rev.\ A {\bf 69}, 062305 (2004).

\bibitem{Fuchs04b}
C.~A. Fuchs and R.~Schack, ``Unknown Quantum States and Operations, a
Bayesian View,'' in {\sl Quantum Estimation Theory}, edited by M.~G.~A.
Paris and J. \v{R}eh\'a\v{c}ek, (Springer-Verlag, Berlin, 2004),
pp.~151--190; {\tt arXiv:quant-ph/0404156}.

\bibitem{Fuchs09a}
C.~A. Fuchs and R.~Schack, ``Priors in Quantum Bayesian Inference,'' in {\sl Foundations of Probability and Physics -- 5}, edited by L.~Accardi et al., AIP Conference Proceedings Vol.~1101, (American Institute of Physics, Melville, NY, 2009), pp.~255--259; {\tt arXiv:0906.1714 [quant-ph]}.

\bibitem{Fuchs11b}
C.~A. Fuchs and R.~Schack, ``A Quantum-Bayesian Route to Quantum-State Space,'' Found.\ Phys. {\bf 41}, 345--356 (2011).

\bibitem{Fuchs12a}
C.~A. Fuchs and R.~Schack, ``Bayesian Conditioning, the Reflection Principle, and Quantum Decoherence,'' in {\sl Probability in Physics}, edited by Y.~Ben-Menahem and M.~Hemmo (Springer, Berlin, Frontiers Collection, 2012), pp.~233--247.

\bibitem{Fuchs13a}
C.~A. Fuchs and R.~Schack, ``Quantum-Bayesian Coherence,'' Rev.\ Mod.\ Phys.\ {\bf 85}, 1693--1715 (2013).

\bibitem{Fuchs14b}
C. A. Fuchs and R. Schack, ``Quantum Measurement and the Paulian Idea,'' in {\sl The Pauli-Jung Conjecture and Its Impact Today}, edited by H. Atmanspacher and C.~A. Fuchs (Imprint Academic, Exeter, UK, 2014), pp.~93--107.

\bibitem{Fuchs15b}
C. A. Fuchs and R. Schack, ``QBism and the Greeks:\ Why a Quantum State Does Not Represent an Element of Physical Reality,'' Physica Scripta {\bf 90}, 015104 (2015).

\bibitem{Penrose90}
R. Penrose, {\sl The Emperor's New Mind:\ Concerning Computers, Minds, and the Laws of Physics}, (Oxford University Press, Oxford, 1990), p.~295.

\bibitem{Timpson08a}
C. G. Timpson, ``Quantum Bayesianism:\ A Study,'' Stud.\ Hist.\ Phil.\ Mod.\ Phys.\ {\bf 39}, 579--609 (2008).

\bibitem{Wheeler71a}
 J.~A. Wheeler, ``Transcending the Law of Conservation of
Leptons,'' in {\sl Atti del Convegno Internazionale sul Tema:\ The
Astrophysical Aspects of the Weak Interaction\/} (Cortona ``Il
Palazzone,'' 10-12 Giugno 1970), Accademia Nationale die Lincei,
Quaderno N.~{\bf 157} (1971), pp.~133--164.

\bibitem{Wheeler73a}
 J.~A. Wheeler, ``From Relativity to Mutability,'' {\sl The
Physicist's Conception of Nature}, edited by J.~Mehra (D.~Reidel,
Dordrecht, 1973), pp.~202--247.

\bibitem{Wheeler74a}
 J.~A. Wheeler, ``From Mendel\'{e}ev's Atom to the Collapsing
Star,'' in {\it Philosophical Foundations of Science}, edited by
R.~J. Seeger and R.~S. Cohen (Reidel, Dordrecht, 1974), pp.~275--301.

\bibitem{Wheeler74b}
 J.~A. Wheeler, ``The Universe as Home for Man,'' Am.\ Sci.\ {\bf
62}, 683--691 (1974).  This is an abridged, early version of
Ref.~\cite{Wheeler75c}.

\bibitem{Wheeler75a}
 J.~A. Wheeler, ``From Magnetic Collapse to Gravitational
Collapse:\ Levels of Understanding Magnetism,'' in {\sl Role of
Magnetic Fields in Physics and Astrophysics}, Ann.\ NY Acad.\ Sci.\
{\bf 257}, edited by V.~Canuto (NY Academy of Sciences, NY, 1975),
pp.~189--221.

\bibitem{Wheeler75b}
 J.~A. Wheeler, ``Another Big Bang?'' Am.\ Sci.\ {\bf 63}, 138
(1975). This is a letter in reply to a reader's comments on
Ref.~\cite{Wheeler74b}.

\bibitem{Wheeler75c}
 J.~A. Wheeler, ``The Universe as Home for Man,'' {\sl The Nature
of Scientific Discovery:\ A Symposium Commemorating the 500th
Anniversary of the Birth of Nicolaus Copernicus}, edited by
O.~Gingerich (Smithsonian Institution Press, City of Washington,
1975), pp.~261--296, discusion pp.~575--587.

\bibitem{PattonWheeler75}
C.~M. Patton and J.~A. Wheeler, ``Is Physics Legislated by
Cosmogony?,'' in {\sl Quantum Gravity:~An Oxford Symposium}, edited
by C.~J. Isham, R.~Penrose, and D.~W. Sciama (Clarendon Press,
Oxford, 1975), pp.~538--605.

\bibitem{Wheeler76a}
 J.~A. Wheeler, ``Include the Observer in the Wave Function?,''
Fundamenta Scientiae:\ Seminaire sur les Fondements des Sciences
(Strasbourg) {\bf 25}, 9--35 (1976).

\bibitem{WheelerPatton77}
J.~A. Wheeler and C.~M. Patton, ``Is Physics Legislated by
Cosmogony?,'' in {\sl Encyclopedia of Ignorance:\ Everything You Ever
Wanted to Know about the Unknown}, edited by R.~Duncan and
M.~Weston-Smith (Pergamon, Oxford, 1977), pp.~19--35. This is an
abridged version of Ref.~\cite{PattonWheeler75}.

\bibitem{Wheeler77a}
 J.~A. Wheeler, ``Genesis and Observership,'' in {\sl Foundational
Problems in the Special Sciences:\ Part Two of the Proceedings of the
Fifth International Congress of Logic, Methodology and Philosophy of
Science, London, Canada -- 1975}, edited by R.~E. Butts and
J.~Hintikka (D.~Riedel, Dordrecht, 1977), pp.~3--33.

\bibitem{Wheeler77b}
 J.~A. Wheeler, ``Include the Observer in the Wave Function?,''
{\sl Quantum Mechanics, a Half Century Later:\ Papers of a Colloquium
on Fifty Years of Quantum Mechanics, Held at the University Louis
Pasteur, Strasbourg, May 2--4, 1974}, edited by J.~Leite Lopes and
M.~Paty (D.~Reidel, Dordrecht, 1977), pp.~1--18.  This is a reprint
of Ref.~\cite{Wheeler76a}.

\bibitem{Wheeler78a}
 J.~A. Wheeler, ``The `Past' and the `Delayed-Choice' Double-Slit
Experiment,'' {\sl Mathematical Foundations of Quantum Theory},
edited by A.~R. Marlow (Academic Press, New York, 1978), pp.~9--48.

\bibitem{Wheeler79a}
 J.~A. Wheeler, ``Parapsychology---A Correction,'' Science {\bf
205}, 144 (1979).  This correction refers to something stated in
Wheeler's talk at the annual meeting of the AAAS (reported later in corrected form in
Ref.~\cite{Wheeler81a}).

\bibitem{Wheeler79b}
 J.~A. Wheeler, ``Frontiers of Time,'' in {\sl Problems in the
Foundations of Physics, Proceedings of the International School of
Physics ``Enrico Fermi,'' Course LXXII}, edited by G.~Toraldo di
Francia (North-Holland, Amsterdam, 1979), pp.~395--492.

\bibitem{Wheeler79c}
 J.~A. Wheeler, ``The Quantum and the Universe,'' in {\sl
Relativity, Quanta, and Cosmology in the Development of the
Scientific Thought of Albert Einstein, Vol.~II}, edited by
F.~de~Finis (Johnson Reprint Corp., New York, 1979), pp.~807--825.

\bibitem{Wheeler79d}
 J.~A. Wheeler, ``The Superluminal,'' New York Review of Books,
27 September 1979, p.~68.

\bibitem{Wheeler79e}
 J.~A. Wheeler, ``Collapse and Quantum as Lock and Key,'' Bull.\
Am.\ Phys.\ Soc., Series II {\bf 24}, 652--653 (1979).

\bibitem{Wheeler80a}
 J.~A. Wheeler, ``Beyond the Black Hole,'' in {\sl Some
Strangeness in the Proportion:\ A Centennial Symposium to Celebrate
the Achievements of Albert Einstein}, edited by H.~Woolf
(Addison-Wesley, Reading, MA, 1980), pp.~341--375, discussion
pp.~381--386.

\bibitem{Wheeler80b}
 J.~A. Wheeler, ``Pregeometry:\ Motivations and Prospects,'' in
{\sl Quantum Theory and Gravitation:\  Proceedings of a Symposium Held
at Loyola University, New Orleans, May 23-26, 1979}, edited by A.~R.
Marlow (Academic Press, New York, 1980), pp.~1--11.

\bibitem{Wheeler80c}
 J.~A. Wheeler, ``Law without Law,'' {\sl Structure in Science and
Art}, edited by P.~Medawar and J.~H. Shelley (Elsevier, Amsterdam,
1980), pp.~132--154.

\bibitem{Wheeler80d}
 J.~A. Wheeler, ``Delayed-Choice Experiments and the
Bohr-Einstein Dialogue,'' in {\sl American Philosophical Society and
the Royal Society:\ Papers Read at a Meeting, June 5, 1980}, (American
Philosophical Society, Philadelphia, 1980), pp.~9--40.

\bibitem{Wheeler81a}
 J.~A. Wheeler, ``Not Consciousness but the Distinction Between
the Probe and the Probed as Central to the Elemental Quantum Act of
Observation,'' in {\sl The Role of Consciousness in the Physical
World}, edited by R.~G. Jahn (Westview Press, Boulder, CO, 1981),
pp.~87--111.

\bibitem{Wheeler81b}
 J.~A. Wheeler, ``The Participatory Universe,'' Science81 {\bf
2}(5), 66--67 (1981).

\bibitem{Wheeler81c}
 J.~A. Wheeler, ``The Elementary Quantum Act as Higgledy-Piggledy
Building Mechanism,'' in {\sl Quantum Theory and the Structures of
Time and Space:\ Papers Presented at a Conference Held in Tutzing,
July, 1980}, edited by L.~Castell and C.~F. von Weizs\"acker (Carl
Hanser, Munich, 1981), pp.~27--30.

\bibitem{Wheeler82a}
 J.~A. Wheeler, ``The Computer and the Universe,'' Int.\ J. Theo.\
Phys.\ {\bf 21}, 557--572 (1982).

\bibitem{Wheeler82b}
 J.~A. Wheeler, ``Particles and Geometry,'' in {\sl Unified
Theories of Elementary Particles:\ Critical Assessment and Prospects},
Lecture Notes in Physics {\bf 160}, edited by P.~Breitenlohner and
H.~P. D\"urr (Springer-Verlag, Berlin, 1982), pp.~189--217.

\bibitem{Wheeler82c}
J.~A. Wheeler, ``Bohr, Einstein, and the Strange Lesson of the
Quantum,'' in {\sl Mind in Nature:\ Nobel Conference XVII, Gustavus
Adolphus College, St.~Peter, Minnesota}, edited by R.~Q. Elvee
(Harper \& Row, San Francisco, CA, 1982), pp.~1--23, and discussions
pp.~23--30, 88--89, 112--113, and 148--149.

\bibitem{Wheeler82d}
 J.~A. Wheeler, ``Physics and Austerity:\ Law without Law,''
University of Texas preprint, 1--87 (1982).

\bibitem{Wheeler82e}
 J.~A. Wheeler, ``Black Holes and New Physics,''
Discovery:\ Research and Scholarship at the University of Texas at
Austin {\bf 7}(2), 4--7 (Winter 1982).

\bibitem{Wheeler83a}
 J.~A. Wheeler, ``On Recognizing `Law without Law':\ Oersted Medal
Response at the Joint APS-AAPT Meeting, New York, 25 January 1983,''
Am.\ J. Phys.\ {\bf 51}, 398--404 (1983).

\bibitem{Wheeler83b}
 J.~A. Wheeler, ``Elementary Quantum Phenomenon as Building Unit,''
in {\sl Quantum Optics, Experimental Gravity, and Measurement Theory},
edited by P.~Meystre and M.~O. Scully (Plenum Press, New York, 1983),
pp.~141--143.

\bibitem{Wheeler83c}
 J.~A. Wheeler, ``Law without Law,'' in {\sl Quantum Theory and
Measurement}, edited by J.~A. Wheeler and W.~H. Zurek (Princeton
University Press, Princeton, 1983), pp.~182--213.

\bibitem{Wheeler83d}
 J.~A. Wheeler, ``Guest Editorial:\ The Universe as Home for
Man,'' in {\sl The Dynamic Universe:\ An Introduction to Astronomy},
edited by T.~P. Snow (West Pub.\ Co., St.~Paul, Minnesota, 1983),
pp.~108--109. This is an excerpt from Ref.~\cite{Wheeler75c}.

\bibitem{Wheeler83e}
 J.~A. Wheeler, ``Physics and Austerity,'' in {\sl Krisis},
Vol.~1, No.~2, edited by I.~Masculescu (Klinckscieck, Paris, 1983),
pp.~671--675.

\bibitem{Wheeler84a}
 J.~A. Wheeler, ``Quantum Gravity:\ The Question of Measurement,''
in {\sl Quantum Theory of Gravity:\ Essays in Honor of the 60th
Birthday of Bryce S. DeWitt}, edited by S.~M. Christensen (Adam
Hilger, Bristol, 1984), pp.~224--233.

\bibitem{Wheeler84c}
 J.~A. Wheeler, ``Bits, Quanta, Meaning,'' in {\sl Problems in
Theoretical Physics}, edited by A.~Giovannini, F.~Mancini, and
M.~Marinaro (University of Salerno Press, Salerno, 1984),
pp.~121--141.

\bibitem{Wheeler84d}
 J.~A. Wheeler, ``Bits, Quanta, Meaning,'' in {\sl Theoretical
Physics Meeting:\ Atti del Convegno, Amalfi, 6-7 Maggio 1983},
(Edizioni Scientifiche Italiene, Naples, 1984), pp.~121--134.

\bibitem{Wheeler85a}
 J.~A. Wheeler, ``Bohr's `Phenomenon' and `Law without Law','' in
{\sl Chaotic Behavior in Quantum Systems:\ Theory and Applications},
edited by G.~Casati (Plenum Press, New York, 1985), pp.~363--378.

\bibitem{Wheeler85b}
 J.~A. Wheeler, ``Niels Bohr, the Man,'' Phys.\ Today {\bf
38}(10), 66--72 (1985).

\bibitem{Wheeler85c}
 J.~A. Wheeler, ``Delayed-Choice Experiments and the
Bohr-Einstein Dialogue,'' in {\sl Niels Bohr:\ A Profile}, edited by
A.~N. Mitra, L.~S. Kothari, V.~Singh, and S.~K. Trehan (Indian
National Science Academy, New Delhi, 1985), pp.~139--168.  This is a
reprint of Ref.~\cite{Wheeler80d}.

\bibitem{Wheeler86a}
 J.~A. Wheeler, ``Hermann Weyl and the Unity of Knowledge,'' Am.\
Sci.\ {\bf 74}, 366--375 (1986).

\bibitem{Wheeler86b}
 J.~A. Wheeler, ``Niels Bohr:\ The Man and his Legacy,'' in {\sl
The Lesson of Quantum Theory}, edited by J.~de~Boer, E.~Dal, and
O.~Ulfbeck (Elsevier, Amsterdam, 1986), pp.~355--367.

\bibitem{Wheeler86c}
 J.~A. Wheeler, ``\,`Physics as Meaning Circuit':\ Three Problems,''
in {\sl Frontiers of Nonequilibrium Statistical Physics}, edited by
G.~T. Moore and M.~O. Scully (Plenum Press, New York, 1986),
pp.~25--32.

\bibitem{Wheeler86d}
J.~A. Wheeler, ``Foreword,'' in J.~D. Barrow and F.~J. Tipler, {\sl
The Anthropic Cosmological Principle}, (Oxford University Press,
Oxford, 1986), pp.~vii--ix.

\bibitem{Wheeler87a}
 J.~A. Wheeler, ``How Come the Quantum'' in {\it New Techniques
and Ideas in Quantum Measurement Theory}, edited by D.~M.
Greenberger, Ann.\ New York Acad.\ Sci.\ {\bf 480}, 304--316 (1987).

\bibitem{Wheeler87b}
 J.~A. Wheeler, ``Foreword'' in H.~Weyl, {\sl The Continuum:\ A
Critical Examination of the Foundation of Analysis}, translated by
S.~Pollard and T.~Bole (Thomas Jefferson University Press,
Kirksville, MO, 1987), pp.~ix--xiii.  This is an excerpt from
Ref.~\cite{Wheeler86a}.

\bibitem{Wheeler88a}
J.~A. Wheeler, ``World as System Self-Synthesized by Quantum
Networking,'' IBM J. Res.\ Develop.\ {\bf 32}, 4--15 (1988).

\bibitem{Wheeler88b}
 J.~A. Wheeler, ``World as System Self-Synthesized by Quantum
Networking,'' in {\sl Probability in the Sciences}, edited by
E.~Agazzi (Kluwer, Dordrecht, 1988), pp.~103--129.  This is a reprint
of Ref.~\cite{Wheeler88a}.

\bibitem{Wheeler88c}
 J.~A. Wheeler, ``Hermann Weyl and the Unity of Knowledge,'' in
{\sl Exact Sciences and their Philosophical Foundations}, edited by
W.~Deppert {\it et al}.~(Lang, Frankfurt am Main, 1988),
pp.~366--375. This is an expanded version of Ref.~\cite{Wheeler86a}.

\bibitem{Wheeler89a}
J.~A. Wheeler, ``Bits, Quanta, Meaning,'' in {\sl Festschrift in
Honour of Eduardo R. Caianiello}, edited by A.~Giovannini, F.~Mancini,
M.~Marinaro, and A.~Rimini (World Scientific, Singapore, 1989),
pp.~133--154.

\bibitem{Wheeler90a}
 J.~A. Wheeler, ``Information, Physics, Quantum:\ the Search for
Links,'' in {\sl Proceedings of the 3rd International Symposium on
Foundations of Quantum Mechanics in the Light of New Technology},
edited by S.~Kobayashi, H.~Ezawa, Y.~Murayama, and S.~Nomura
(Physical Society of Japan, Tokyo, 1990), pp.~354--368.

\bibitem{Wheeler92a}
 J.~A. Wheeler, ``Recent Thinking about the Nature of the Physical
World:\ It from Bit,'' in {\sl Frontiers in Cosmic Physics:\ Symposium
in Memory of Serge Alexander Korff}, Ann.\ NY Acad.\ Sci.\ {\bf 655},
edited by R.~B. Mendell and A.~I. Mincer (NY Academy of Sciences, NY,
1992), pp.~349--364.

\bibitem{Wheeler92b}
 J.~A. Wheeler, {\sl At Home in the Universe}, (American Institute
of Physics Publishing, New York, 1992).

\bibitem{Wheeler94a}
 J.~A. Wheeler, ``Time Today,'' in {\sl Physical Origins of Time
Asymmetry}, edited by J.~J. Halliwell, J.~P\'erez-Mercader, and W.~H.
Zurek (Cambridge University Press, Cambridge, 1994), pp.~1--29.

\bibitem{Wheeler98a}
J.~A. Wheeler (with K.~W. Ford), {\sl  Geons, Black Holes, and Quantum
Foam:\ A Life in Physics}, (W.~W. Norton, New York, 1998).

\bibitem{Wheeler98b}
J.~A. Wheeler, letter to Carroll Alley, dated 13 March 1998.

\bq
\indent
I want you and Einstein to jolt the world of physics into an
understanding of the quantum because the quantum surely
contains---when unraveled---the most wonderful insight we could ever
hope to have on how this world operates, something equivalent in
scope and power to the greatest discovery that science has ever yet
yielded up: Darwin's Evolution.

You know how Einstein wrote to his friend in 1908, ``This quantum
business is so incredibly important and difficult that everyone should
busy himself with it.'' \ldots\ Expecting something great when two
great minds meet who have different outlooks, all of us in this
Princeton community expected something great to come out from Bohr and
Einstein arguing the great question day after day---the central
purpose of Bohr's four-month, spring 1939 visit to Princeton---I, now,
looking back on those days, have a terrible conscience because the
day-after-day arguing of Bohr was not with Einstein about the quantum
but with me about the fission of uranium. How recover, I ask myself
over and over, the pent up promise of those long-past days? Today, the
physics community is bigger and knows more than it did in 1939, but it
lacks the same feeling of {\bf desperate} puzzlement. I want to
recapture that feeling for us all, even if it is my last act on Earth.
\eq

\bibitem{Wheeler98c}
J.~A. Wheeler, ``The Eye and the U,'' letter dated 19 March 1998 to
Abner Shimony, with carbon copies to Arthur Wightman, Peter Mayer,
Demetrios Christodoulou, Larry Thomas, Elliot Lieb, Charles Misner,
Frank Wilczek, Kip Thorne, Ben Schumacher, William Wootters, Jim
Hartle, Edwin Taylor, Ken Ford, Freeman Dyson, Peter Cziffra, and
Arkady Plotnitsky.

\bibitem{Wheeler99a}
J.~A. Wheeler, ``Information, Physics, Quantum:\ The Search for
Links,'' in {\sl Feynman and Computation:\ Exploring the Limits of
Computers}, edited by A.~J.~G. Hey (Perseus Books, Reading, MA,
1999), pp.~309--336.

\bibitem{Wheeler00}
J.~A. Wheeler, ``\,`A Practical Tool,' But Puzzling, Too,'' New York
Times, 12 December 2000.

\bibitem{Bub16a}
J. Bub, {\sl Bananaworld:\ Quantum Mechanics for Primates}, (Oxford University Press, Oxford, 2016), p.~232.

\bibitem{Ladyman07a}
J. Ladyman, D. Ross, D. Spurrett, and J. Collier, {\sl Everything Must Go:\ Metaphysics Naturalized}, (Oxford University Press, Oxford, 2007), p.~184.

\bibitem{Samizdat1}
C.~A. Fuchs, {\sl Coming of Age with Quantum Information:\ Notes on a Paulian Idea}, (Cambridge University Press, Cambridge, UK, 2010).

\bibitem{Samizdat2}
C.~A. Fuchs, {\sl My Struggles with the Block Universe:\ Selected Correspondence, January 2001 -- May 2011}, edited by Blake C. Stacey, foreword by Maximilian Schlosshauer (2014), 2,349 pages; {\tt arXiv:1405.2390 [quant-ph]}.

\bibitem{Cabello15a}
A. Cabello, ``Interpretations of Quantum Theory:\ A Map of Madness,'' {\tt arXiv:1509.04711 [quant-ph]} (2015).

\bibitem{Einstein49}
A. Einstein, ``Remarks Concerning the Essays Brought Together in This Co-operative Volume,'' in {\sl Albert Einstein:\ Philosopher-Scientist}, edited by P.~A. Schilpp, (Tudor Publishing Co., New York, 1949), pp.~665--688.

\bibitem{Brukner01}
\v{C}. Brukner and A. Zeilinger, ``Conceptual Inadequacy of the Shannon Information in Quantum Measurements,'' Phys.\ Rev. A, {\bf 63}, 022113 (2001).

\bibitem{Kofler10}
J. Kofler and A. Zeilinger, ``Quantum Information and Randomness,'' Europ.\ Rev.\ {\bf 18}, 469--480 (2010).

\bibitem{Zeilinger05a}
A. Zeilinger, ``The Message of the Quantum,'' Nature {\bf 438}, 743 (2005).

\end{thebibliography}
\end{document}